\documentclass[twocolumn,showpacs,preprintnumbers,amsmath,amssymb]{revtex4}
\usepackage{graphicx}
\begin{document}
\title{Role of the Mean-field in Bloch Oscillations of a Bose-Einstein Condensate in an Optical Lattice and Harmonic Trap}

\author{R. Zhang, R. E. Sapiro, R. R. Mhaskar, G. Raithel}

\affiliation{FOCUS Center, Department of Physics, University of
Michigan, Ann Arbor, MI 48109}

\date{\today}

\begin{abstract}

Using the Crank-Nicholson method, we study the evolution of a Bose-Einstein
condensate in an optical lattice and harmonic trap. The condensate is
excited by displacing it from the center of the harmonic trap. The mean
field plays an important role in the Bloch-like oscillations that occur
after sufficiently large initial displacement. We find that a moderate mean
field significantly suppresses the dispersion of the condensate in momentum
space. When the mean field becomes large, soliton and vortex structures
appear in the condensate wavefunction.

\end{abstract}

\maketitle

\section{Introduction}

The dynamics of Bose-Einstein condensates (BECs) in optical lattices have
been studied using both theoretical and experimental methods (for a review,
see Ref.~\cite{morsch1}). In many cases, the atomic interaction cannot be
ignored due to the combination of low temperature and high density found in
a BEC. This interaction is often accounted for as a mean-field effect. The
role of the mean field in optical lattices has received theoretical
attention in a number of phenomena, such as pulsed atom lasing in an optical
lattice~\cite{chiofalo,Trom}, number squeezing of a BEC in an optical
lattice~\cite{McKagan}, Landau-Zener tunneling of a
BEC~\cite{Holthaus,Jona}, and the dynamics of a BEC in an accelerating
optical lattice~\cite{choi}. In related experimental work,
mean-field-induced four-wave mixing of matter waves has been
observed~\cite{deng}, and the mean field has been shown to change the
Landau-Zener tunneling rate~\cite{Jona,Cristiani}. In this paper, we
simulate the dynamics of a BEC in a combined potential consisting of an
optical lattice and a harmonic trap. The initial displacement of the BEC
from the center of the harmonic trap triggers motion similar to a Bloch
oscillation. Bloch oscillations of BECs in optical lattices have been
studied theoretically~\cite{choi,berg,Scott} and observed
experimentally~\cite{anderson,morsch,li,gustavsson,fattori,zhang} in various
configurations. In this paper, we focus on the systematic investigation of
the role of the mean field in Bloch oscillations in harmonic potentials. We
find that the stability of the BEC during the oscillations depends on the
interplay between the inhomogeneity of the applied force and the mean field.
In recent experiments on Bloch oscillations driven by a constant external
force, the mean field was shown to cause momentum dispersion of the
BEC~\cite{gustavsson,fattori}. Interestingly, in Bloch oscillations in an
approximately harmonic magnetic trap, this dispersion can be suppressed by
the mean field~\cite{zhang}. All of these findings are consistently
described by the results in this paper.

The paper is organized as follows. In Sec.~\ref{sec:GPE}, we discuss the
methods and formalism used in this work. In Sec.~\ref{sec:dispersion}, we
explore the effect of the mean field on dispersion characteristics in an
inhomogeneous force field. In Sec.~\ref{sec:soliton}, we discuss solitions
and vortices observed in the case of a large mean field. In
Sec.~\ref{sec:conclusion}, we conclude the paper.

\section{Integrating the Gross-Pitaevskii Equation}
\label{sec:GPE}
At zero temperature and in the limit of weak excitation, the
BEC wavefunction, $\Psi(\textbf{r},t)$, can be described by the
Gross-Pitaevskii (GP) equation~\cite{Gross, Pitaevskii},
\begin{equation}
i\hbar \frac{\partial\Psi(\textbf{r},t)}{\partial
t}=[-\frac{\hbar^2}{2M}\nabla_{\rm{\textbf{r}}}^2 +
V_{\rm{ext}}(\textbf{r})+
NV_{\rm{int}}|\Psi(\textbf{r},t)|^2]\Psi(\textbf{r},t) \,,\label{gp}
\end{equation}
where $M$ is the atomic mass, $N$ is the total atom number,
$V_{\rm{ext}}(\textbf{r})$ is the external potential and $V_{\rm{int}}$
characterizes the strength of the mean-field interaction between the atoms,
defined as $V_{\rm{int}}\equiv4\pi \hbar^2a/M$, with $a$ being the s-wave
scattering length of the atom.

We assume that the BEC is confined in a cylindrically symmetric harmonic
trap, which can be written as $V_{\rm{trap}}(\textbf{r})=
\frac{1}{2}M(\omega_{\rm{r}}^2r^2+\omega_{\rm{z}}^2z^2)$ using cylindrical
coordinates $\textbf{r}=(r,\theta,z)$. In addition, an optical lattice is
applied to the BEC in the direction of the axis of symmetry,
$V_{\rm{OL}}(\textbf{r})=V_0\sin^2(2\pi z/\lambda)$, where $\lambda$ is the
wavelength of the lattice laser. In the case of zero angular momentum about
the symmetry axis, the GP equation can be rewritten in cylindrical
coordinates as

\begin{align}
i\hbar \frac{\partial\Psi(r,z,t)}{\partial t}&= [
-\frac{\hbar^2}{2M}(\frac{1}{r}\frac{\partial}{\partial
r}r\frac{\partial}{\partial r}+\frac{\partial ^2}{\partial z^2}) \nonumber
\\ &+ \frac{1}{2}M(\omega_{\rm{r}}^2r^2+
\omega_{\rm{z}}^2z^2)+V_{\rm{0}}\sin^2(\frac{2\pi z}{\lambda}) \nonumber
\\ & +NV_{\rm{int}}|\Psi(r,z,t)|^2 ] \Psi(r,z,t) \, .\label{gp1}
\end{align}
The normalization condition is given by
\begin{equation}
\int 2\pi r|\Psi(r,z,t)|^2 drdz= 1 \,.\label{no}
\end{equation}
In this paper, we numerically solve this 2-dimensional (2D) GP equation to
study the effect of the mean field, $NV_{\rm{int}}|\Psi(r,z,t)|^2$, on the
dynamics of the BEC in the optical lattice.

A well-established numerical method for solving the GP equation is the
Crank-Nicholson algorithm~\cite{crank}. To arrive at boundary conditions
that are conducive to solving the problem using the Crank-Nicholson method,
we introduce a wavefunction $\varphi(r,z,t)$, defined as
\begin{equation}
\frac{\varphi(r,z,t)}{r}=\Psi(r,z,t)\,.
\end{equation}
The boundary conditions for $\varphi(r,z,t)$ are that
$\varphi(r,z,t)\rightarrow $~0 when $r \rightarrow$~0 or $\infty$, or when
$|z| \rightarrow \infty$. The GP equation in term of $\varphi(r,z,t)$
becomes
\begin{align}
i\hbar \frac{\partial\varphi(r,z,t)}{\partial t}&=
\left[-\frac{\hbar^2}{2M}(\frac{\partial ^2}{\partial
r^2}+\frac{1}{r^2}-\frac{1}{r}\frac{\partial}{\partial r}+\frac{\partial
^2}{\partial z^2}) \right. \nonumber \\&+ \frac{1}{2}M(\omega_{\rm{r}}^2r^2+
\omega_{\rm{z}}^2z^2) +V_{\rm{0}}\sin^2(\frac{2\pi z}{\lambda}) \nonumber
\\ &\left.+NV_{\rm{int}}|\frac{\varphi(r,z,t)}{r}|^2 \right]\varphi(r,z,t) \,.\label{gp2}
\end{align}

To reduce the 2D GP equation into tractable 1D steps, it is split into
separate radial and axial equations:
\begin{align}
i\hbar \frac{\partial\varphi(r,z,t)}{\partial t}&=
\left[-\frac{\hbar^2}{2M}(\frac{\partial ^2}{\partial
r^2}+\frac{1}{r^2}-\frac{1}{r}\frac{\partial}{\partial r})+
\frac{1}{2}M\omega_{\rm{r}}^2r^2 \right.\nonumber
\\&\left.+\frac{1}{2}NV_{\rm{int}}|\frac{\varphi(r,z,t)}{r}|^2 \right]\varphi(r,z,t)\label{gp3}
\\
i\hbar \frac{\partial\varphi(r,z,t)}{\partial t}&=
\left[-\frac{\hbar^2}{2M}\frac{\partial ^2}{\partial
z^2}+\frac{1}{2}M\omega_{\rm{z}}^2z^2+V_{\rm{0}}\sin^2(\frac{2\pi
z}{\lambda})\right.\nonumber
\\&\left.+\frac{1}{2}NV_{\rm{int}}|\frac{\varphi(r,z,t)}{r}|^2 \right]\varphi(r,z,t)\,.\label{gp4}
\end{align}
In each time increment of the integration, the evolution of $\varphi(r,z,t)$
is broken up into two steps. In the first step, we fix the $z$ coordinate
and solve Eq.~\ref{gp3} using the standard 1D Crank-Nicholson method. In the
second step, the resultant $\varphi(r,z,t)$ is entered into Eq.~\ref{gp4} as
the initial wavefunction to calculate the time evolution in the
$z$-direction while the $r$ coordinate is fixed, again using the 1D
Crank-Nicholson method. The time and spatial increments used in our
simulations are 100~ns and 50~nm, respectively. A detailed discussion of
this variation of the Crank-Nicholson method for solving the 2D GP equation
can be found in Refs.~\cite{edwards,adhikari1,adhikari}.

\section{Dispersion of the BEC during Bloch Oscillations}
\label{sec:dispersion}

\begin{figure}[!]
\centerline{ \scalebox{.52} {\includegraphics{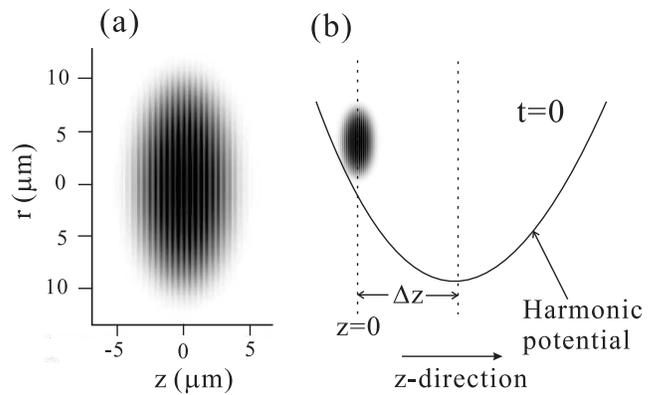}}} \caption{(a)
Initial BEC wavefunction for 25000 atoms (linear gray-scale representation).
(b) Relative position of the BEC with respect to the harmonic trap along the
$z$-direction at $t$=0. } \label{iw}
\end{figure}

In close analogy with our recent experiment~\cite{zhang}, we consider a BEC
with a fixed number of $^{87}$Rb atoms confined in a cylindrically symmetric
harmonic trap, with trap frequencies $\omega_{\rm{r}}=2\pi\times 20$~Hz and
$\omega_{\rm{z}}=2\pi\times 50$~Hz in the radial and axial directions,
respectively. In the simulations, the atoms are adiabatically loaded into an
optical lattice with a depth of one recoil energy, formed by two
counter-propagating laser beams in the $z$-direction with wavelength
$\lambda$=~852~nm. Then, at time $t=$~0, the harmonic trap is suddenly
shifted in the $z$-direction. To obtain the wavefunction immediately before
the shift, we start with the BEC wavefunction in the 3D harmonic potential
without optical lattice in the absence of a mean field, given by the
well-known ground state of the harmonic potential. The ground state of the
BEC including optical lattice and mean field is then obtained by gradually
ramping up the optical-lattice and mean-field potentials to their full
strengths over an interval of 100~ms, using the formalism described in
Eqs.~(\ref{gp3}) and (\ref{gp4}). The initial BEC wavefunction for 25000
atoms is shown in Fig.~\ref{iw} (a).

After obtaining the initial wavefunction, the magnetic trap is shifted at
time $t=$~0 in the $z$-direction by a distance of $\Delta z$, leading to the
initial condition shown in Fig.~\ref{iw}~(b). The resulting BEC motion
depends on $\Delta z$, and can be broken into several regimes. If $\Delta z$
is small enough that the BEC never gains enough momentum to come close to
the edge of the first Brillouin zone, the BEC performs small oscillations
about the minimum of the harmonic trap. For large $\Delta z$, the BEC
periodically traverses the entire first Brillouin zone. The BEC motion is
similar to a Bloch oscillation, except that the BEC experiences an
inhomogeneous rather than a constant force. The critical displacement,
$\Delta z_{\rm{cr}}$, for Bragg reflection and Bloch oscillations to occur
follows from the requirement that the BEC must reach at least one recoil
energy, $E_{\rm{R}}$, when passing through the minimum of the harmonic trap
(with no lattice). Under our conditions,
$E_{\rm{R}}=h^2/(2M\lambda^2)=h\times3.16$~kHz and $\Delta
z_{\rm{cr}}=\sqrt{2E_{\rm{R}}/(M \omega_z ^2)}=$~17.1~$\mu$m. There is also
an intermediate regime in which only a portion of the BEC undergoes Bragg
reflection. In this paper, we focus on the dynamics of the BEC when $\Delta
z > \Delta z_{\rm{cr}}$ and study how the mean field affects the BEC during
Bloch oscillations.

\begin{figure}[ht]
\centerline{ \scalebox{.5} {\includegraphics{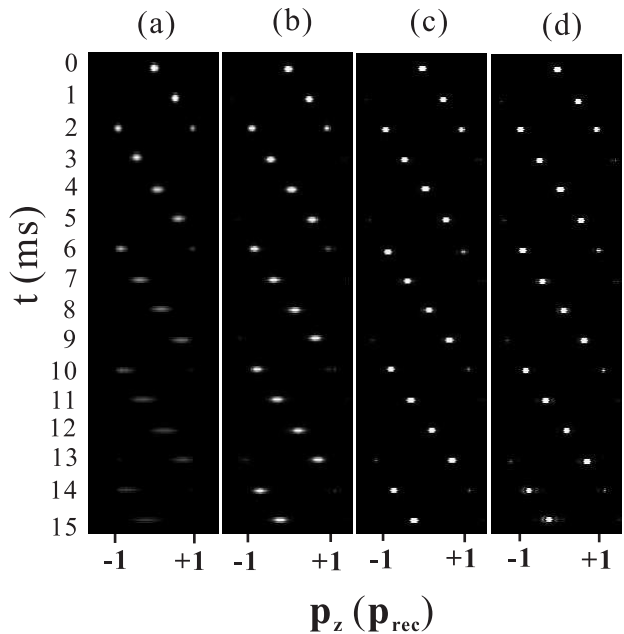}}} \caption{ Momentum
distributions of the BEC as a function of time for different atom numbers,
$N$. (a)-(d) correspond to $N$=0, 2000, 25000, and 60000, respectively. The
brightness represents the atomic density on a linear gray-scale.}
\label{bloch}
\end{figure}

The initial displacement $\Delta z$ is set to be 30~$\mu$m. To visualize the
BEC dynamics, we perform a Fourier transform~\cite{Sekatskii} on the
wavefunction $\Psi(r,z,t)$ and plot the momentum distribution of the BEC as
a function of time. Four different atom numbers are used in the simulations.
The results are shown in Fig.~\ref{bloch} (a)-(d) for $N$=0 (experimentally,
this would be equivalent to a vanishing scattering length), 2000, 25000, and
60000, respectively. The radial (axial) momentum spread of the BEC is given
by the vertical (horizontal) width of the spots in Fig.~\ref{bloch}. The
momentum plots at different times $t$ are stacked in the vertical direction.
In the following discussion, we focus on the $z$-degree of freedom, in which
most of the interesting dynamics occur. As seen in Fig.~\ref{bloch}, the BEC
periodically scans the first Brillouin zone and is Bragg-reflected at times
when it reaches one recoil momentum, $p_{\rm{rec}}$. Since the BEC
experiences a near-constant force during the oscillations, its momentum
varies at a near-constant rate and the overall motion resembles that of a
Bloch oscillation.

\begin{figure}[t]
\centerline{ \scalebox{.5} {\includegraphics{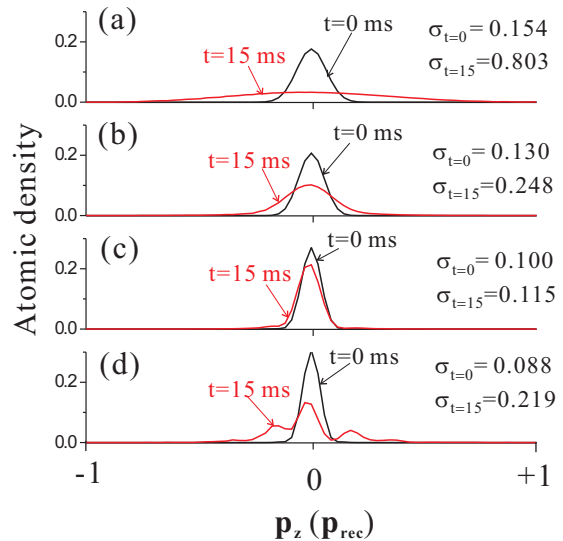}}} \caption{
$z$-momentum probability distribution at t=0 and t=15~ms for $N$=0, 2000,
25000, and 60000, respectively. The center of the $p_{\rm{z}}$ distribution
at t=15~ms is shifted to $p_{\rm{z}}=0$. The values of $\sigma$ are given in
units of $p_{\rm{rec}}$.} \label{mw}
\end{figure}

Comparing the Bloch oscillations for different atom numbers, shown in
Fig.~\ref{bloch} (a)-(d), we notice that without the mean field ($N$=0 in
Fig.~\ref{bloch} (a)), the BEC displays a significantly larger amount of
momentum dispersion than in the cases when the mean field is included. While
the mean field with $N= 25000$, shown in Fig.~\ref{bloch} (c), seems to
greatly suppress this momentum dispersion, mean-field-induced structure
starts to appear in momentum space for $N= 60000$ in Fig.~\ref{bloch} (d).

To better view the momentum dispersion, we plot the atomic density as a
function of $p_{\rm{z}}$ after integrating over $p_{\rm{r}}$, the momentum
in the $r$-direction. The results are shown in Fig.~\ref{mw} (a)-(d),
corresponding to t=0~ms and 15~ms in Fig.~\ref{bloch} (a)-(d), respectively.
We evaluate the full width half maximum, $\sigma$, of each curve using a
Gaussian fit. In Fig.~\ref{mw} (a), with $N=$~0, $\sigma$ increases by a
factor of more than five from t=0~ms to t=15~ms. This dispersion is due to
the fact that the force across the condensate is not uniform. Atoms farther
away from the center of the harmonic trap experience a larger force during
the oscillation, thus reaching $p_{\rm{rec}}$ earlier than atoms closer to
the center. This effect initiates a breathing motion of the BEC wavepacket
in coordinate space and causes the momentum of the BEC to spread as well.
This explanation is supported by an additional simulation in which, for
$N=0$, the $z$-component of the harmonic trap is replaced by a constant
force. In this case, the BEC wavepacket does not disperse during Bloch
oscillations in either coordinate or momentum spaces. The latter result is
confirmed by recent experiments elsewhere~\cite{gustavsson,fattori}, in
which more than 10000 Bloch oscillations are observed when the BEC
experiences a constant external force and the atomic interaction is tuned to
0 through a Feshbach resonance.

\begin{figure}[t]
\centerline{ \scalebox{.55} {\includegraphics{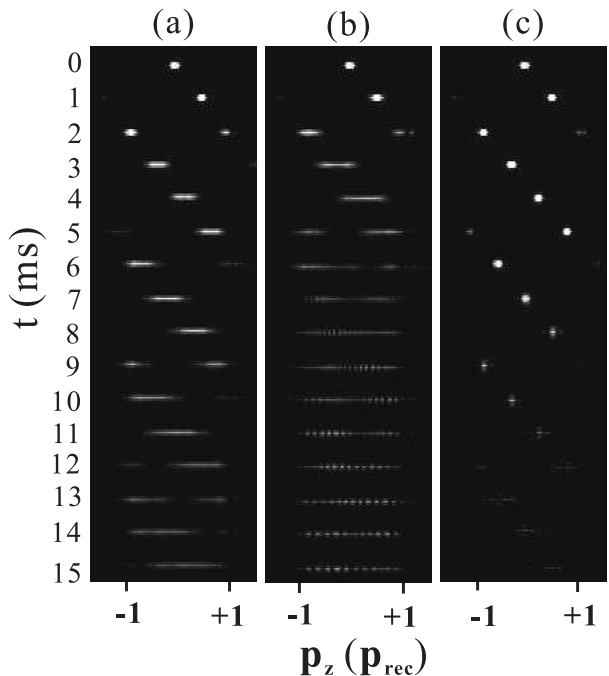}}} \caption{ (a) Same
plot as Fig.~\ref{bloch} for $N=25000$, with a constant external force
applied to the BEC. (b) The harmonic potential in the $z$-direction is first
shifted by $\Delta z = -30~\mu$m, and then switched to an anti-trap. (c) In
addition to (b), the harmonic potential in the $r$-direction is also
switched to an anti-trap and the interaction between condensate atoms is
changed to be attractive.} \label{blochforce}
\end{figure}

The momentum dispersion is greatly suppressed when a moderate mean field is
included in the simulation. For $N$=2000, shown in Fig.~\ref{mw} (b),
$\sigma$ increases by a factor of less than two over 15~ms. When $N$ reaches
25000, there is no significant momentum dispersion over 15~ms, as can be
seen in Fig.~\ref{mw} (c). To qualitatively explain these results, we
consider the mean field potential $V_{\rm{mf}}$ within the Thomas-Fermi
approximation. Before the shift, the BEC has a spatially constant chemical
potential, $\mu$, given by
\begin{equation}
\mu=V_{\rm{mf}}+\frac{1}{2}M(\omega_{\rm{z}}^2z^2+\omega_{\rm{r}}^2r^2)\,.
\end{equation}
Therefore,
$V_{\rm{mf}}=\mu-\frac{1}{2}M(\omega_{\rm{z}}^2z^2+\omega_{\rm{r}}^2r^2)$
inside the BEC. Immediately after the shift, as shown in Fig.~\ref{iw}~(b),
the BEC experiences a total potential $V_{\rm{tot}}$
\begin{equation}
V_{\rm{tot}}=\mu+\frac{1}{2}M\omega_{\rm{z}}^2\Delta
z^2-M\omega_{\rm{z}}^2z\Delta z\,.\label{pot}
\end{equation}
The total force on the BEC, $F_{\rm{tot}}$, is given by
\begin{equation}
{\bf{F}}_{\rm{tot}}=-\nabla V_{\rm{tot}}=M\omega_{\rm{z}}^2\Delta z \,
\hat{\bf {z}}\,.\label{force}
\end{equation}
According to Eq.~\ref{force}, with the mean field included, the BEC
experiences a constant total force when displaced from the center of the
harmonic trap. In this way, the mean field helps stabilize the BEC
wavepacket during the Bloch oscillations. The validity of the Thomas-Fermi
approximation requires the mean-field potential to dominate the external
potential. In our case, the approximation is valid for $N=25000$ but not for
$N=2000$.

\begin{figure}[ht]
\centerline{ \scalebox{.5} {\includegraphics{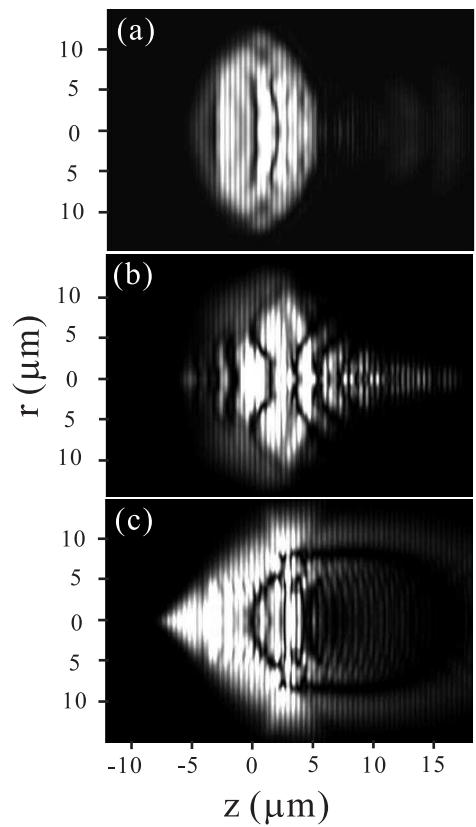}}} \caption{ BEC
spatial density at (a) 15~ms after a 30~$\mu$m displacement, (b) 14~ms after
a 20~$\mu$m displacement, and (c) 10~ms after a 17.2~$\mu$m displacement.
The atom number $N$=60000 in all cases. (Linear gray-scale representation.)}
\label{wave}
\end{figure}

Based on the above argument, when a constant external force is applied to
the BEC, the mean field will cause momentum dispersion. To test this, we
simulate Bloch oscillations under a constant force of the same magnitude as
that caused by a $\Delta z=30$~$\mu m$ shift of the trap. The corresponding
simulation results are shown in Fig.~\ref{blochforce} (a), where momentum
dispersion is clearly seen. This has also been observed experimentally
elsewhere~\cite{gustavsson,fattori}. To expand on our argument, we simulate
the dynamics of the BEC under two other conditions with the same initial
wavefunction as before. In the first case, at t=0 the harmonic potential in
the $z$-direction is first shifted by $\Delta z = -30~\mu$m and then
switched to an anti-trap. Since the mean field now enhances the
inhomogeneity of the effective force acting on the BEC, we expect faster
momentum dispersion, which is verified by simulation results presented in
Fig.~\ref{blochforce} (b). The second case has the same initial conditions
as the first, except that the harmonic potential in the $r$-direction is
also changed to an anti-trap and the atomic interaction is made attractive.
In this case, the momentum dispersion is expected to be suppressed again
because the total force due to the harmonic potential and mean field is
again constant. In Fig.~\ref{blochforce} (c), no momentum dispersion is seen
over several Bloch oscillations, as anticipated. Eventually, the BEC
oscillation becomes unstable due to the intrinsic instability of BECs with
negative scattering length.

\section{Creation of Solitons and Vortices}
\label{sec:soliton}

For the conditions studied in this paper, an increase of the mean field
beyond values corresponding to $N\sim  25000$ causes discrete peaks to
appear in the BEC momentum distribution, as seen in Fig.~\ref{bloch} (d).
This indicates that the internal structure of the BEC wavefunction has been
disturbed. We plot the corresponding BEC density,
$|\Psi(r,z,t=15~\rm{ms})|^2$, in Fig.~\ref{wave} (a), and can see dark
stripes within the BEC. These dark stripes are solitons created by the BEC
standing wave formed during Bragg reflection~\cite{Scott}. Alternatively,
considering the structure of the lowest energy band of the optical lattice,
the soliton formation can be attributed to the anomalous
dispersion~\cite{Eiermann} or negative effective
mass~\cite{Ostrovskaya,Ostrovskaya1} of the BEC near the band edge. The
solitons decay into vortices as the oscillation goes on. The creation of
solitons and vortices during Bloch oscillations was first reported in
Ref.~\cite{Scott}. In this section, we show that the initial displacement
$\Delta z$ has a considerable impact on this phenomenon.
\begin{figure}[t]
\centerline{ \scalebox{.65} {\includegraphics{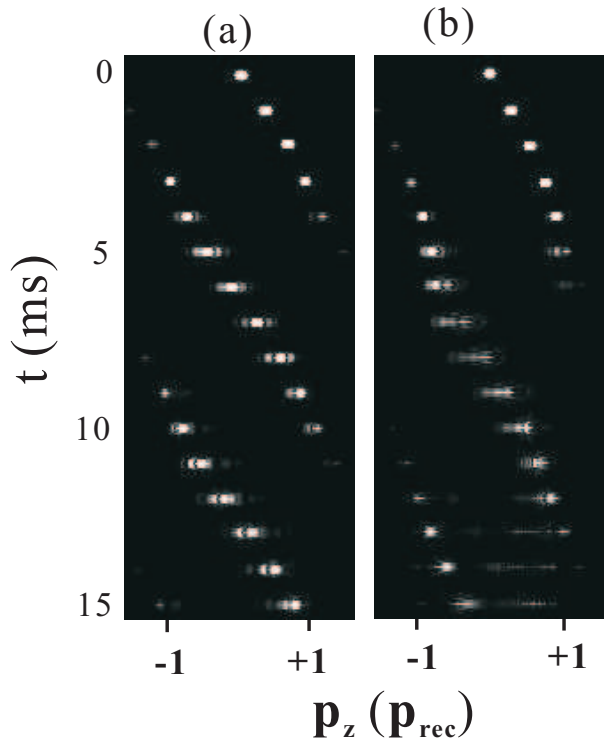}}} \caption{ Same
plot as Fig.~\ref{bloch} for $N=60,000$, with (a) $\Delta z = 20~\mu$m and
(b) $\Delta z = 17.2~\mu$m. Time interval between adjacent images is 1~ms.}
\label{bloch1}
\end{figure}

Figures~\ref{bloch1} (a) and (b) show the BEC momentum distributions for
$\Delta z =20~\mu$m and $\Delta z = 17.2~\mu$m, respectively, both with
$N$=60000. As shown in Fig.~\ref{bloch1} (a), the BEC momentum starts to
develop structure along the $z$-direction after a single Bragg reflection.
In Fig.~\ref{bloch1} (b), a single Bragg reflection causes even more
structure in the BEC momentum distribution. Additionally taking into account
the results shown in Fig.~\ref{bloch} (d), we conclude that a smaller
$\Delta z$ leads to a faster creation of solitons and vortices. This
observation can be explained as follows. Since it is the standing wave
formed during the Bragg reflection that leads to the creation of solitons
and vortices, it is reasonable to assume that the standing wave duration,
$\Delta T$, facilitates the soliton formation. The duration $\Delta T$ can
also be thought of as the Bragg reflection time, which is determined by how
fast the BEC passes through the anti-crossing region between the two lowest
energy bands of the optical lattice~\cite{zhang}. Thus, $\Delta T$ is
proportional to $1/F$, where $F$ is the force applied to the BEC.  Smaller
displacements $\Delta z$ (under the condition $\Delta z >\Delta
z_{\rm{cr}}$) lead to smaller $F$ and larger $\Delta T$, which causes the
solitons and vortices to appear earlier, as observed in the simulations.

\begin{figure}[ht]
\centerline{ \scalebox{.65} {\includegraphics{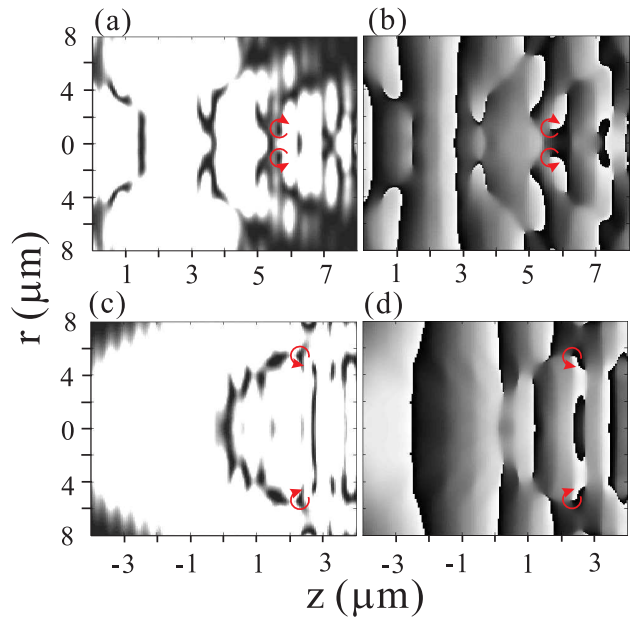}}} \caption{ (Color
online) (a) and (c) are zoomed-in images of Fig.~\ref{wave} (b) and (c),
respectively. (b) and (d) are phase plots corresponding to (a) and (c).
(black=0, white=2$\pi$) } \label{phase}
\end{figure}

Around 14~ms in Fig.~\ref{bloch1} (a) and 10~ms in Fig.~\ref{bloch1} (b),
the solitons start to decay into vortices. The corresponding BEC densities
in coordinate space are shown in Figs.~\ref{wave} (b) and (c), with
zoomed-in images presented in Figs.~\ref{phase} (a) and (c), respectively.
To verify the vortex formation, we plot the phase of the wavefunction in
Figs.~\ref{phase} (b) and (d). Two of the vortices in each of
Figs.~\ref{phase} (b) and (d) are highlighted by circles, where the phase
changes from 0 to 2$\pi$ in one round trip. The arrow represents the
direction of the vortex. Since the vertical axis corresponds to the radial
direction, and the system is azimuthally symmetric, the two highlighted
vortices are part of a single annular vortex ring around $r=0$.

\section{Conclusion}
\label{sec:conclusion}

In this paper, we have simulated the dynamics of a BEC in a combined
potential consisting of a magnetic trap and an optical lattice. After a
sufficiently large initial displacement from the center of the harmonic
trap, the BEC undergoes an oscillatory motion similar to a Bloch
oscillation.  We find that the mean field plays an interesting role during
these oscillations. While a moderate amount of mean field suppresses the
momentum dispersion of the BEC, at large values it causes solitons and
vortices to appear. We have also investigated the dependence of vortex
formation on the initial displacement $\Delta z$. Some of the results
presented in Sec.~\ref{sec:dispersion} have already been verified in a
series of Bloch-oscillation experiments~\cite{gustavsson,fattori,zhang}.
Quantitative studies of the mean-field-induced dispersion characteristics of
Bloch oscillations in harmonic traps may be performed by time-of-flight
measurements of the momentum distribution vs atom number. It will be
challenging to experimentally probe the formation of annular vortices in 2D
Cartesian projections of the BEC obtained using standard time-of-flight
shadow imaging. However, it may be possible to obtain indirect evidence,
such as complex structure in momentum distributions.

This work is supported by AFOSR (grant No. FA9550-07-1-0412) and FOCUS
(NSF grant No. PHY-0114336).

\end{document}